\documentclass[reprint, amsmath,amssymb, aps]{revtex4-2}
\usepackage{graphicx}
\usepackage{xcolor}
\usepackage{float}
\usepackage{dcolumn}
\usepackage{bm}
\usepackage{natbib}
\usepackage{hyperref}

\begin{document}

\preprint{APS/123-QED}

\title{Surface Gravity of Dynamical Horizons: A Causal Perspective}
\author{Anamika Avinash Pathak}
\email{p20190459@hyderabad.bits-pilani.ac.in}
\author{Konka Raviteja}
\email{konka.raviteja@gmail.com}
\author{Swastik Bhattacharya}
\email{swastik@hyderabad.bits-pilani.ac.in}
\author{Sashideep Gutti}
\email{sashideep@hyderabad.bits-pilani.ac.in}

\affiliation{Department of Physics, Birla Institute of Technology and Sciences-Pilani \\Hyderabad, 500078, India}

\date{\today} 

\begin{abstract}
We consider marginally trapped surfaces in a spherically symmetric spacetime evolving due to the presence of a  perfect fluid in D-dimensions and look at the various definitions of the surface gravity for these marginally trapped surfaces. We show that using Einstein equations it is possible to simplify and obtain general formulae for the surface gravity in terms of invariant quantities defined at these marginally trapped surfaces like area radius, cosmological constant and principal values of the energy-momentum tensor $\rho, p$. We then correlate these expressions of surface gravity to the cases of dynamical horizons and timelike tubes and find which proposals of surface gravity are causally sensitive as these surfaces undergo causal transitions from spacelike to timelike and vice versa.
\end{abstract}

\maketitle

\section{Introduction} \label{sec:intro}

Black holes are among the most mysterious objects that exist in our universe. The formation of black holes, their evolution, and their mergers are fields of intense study over many decades.  One intriguing feature of black holes  is the relation between gravity and thermodynamics. The event horizon of the black hole is found to possess entropy and temperature. The origin of the entropy of the black hole and its description in terms of microstates is yet to be properly understood. The connection between thermodynamics and black holes is a well-established area of physics. A relatively less understood phenomenon is the thermodynamics of a black hole that is in the process of evolution. \\

Black hole thermodynamics has been an area of intense study and analysis since the discovery of black hole spacetimes. The connection between the surface gravity of a stationary/static black hole event horizon and temperature is very well established. The relation between black hole entropy and its area has been firmly placed on a strong theoretical foundation due to the presence of Hawking radiation. Though the non-evolving black hole thermodynamics is very well understood, the realistic scenario involving an evolving black hole is still at the initial stages of its formulation. The main reason for this is due to the fact that the dynamical phenomenon describing the formation and evolution of a black hole is extremely complicated, that is except for highly special situations, obtaining analytically solvable solutions in general relativity is difficult.
\\

To capture the features of evolving horizons and trapped regions, Ashtekar et al. \cite{ashtekar2002dynamical,ashtekar2004isolated} defined dynamical horizons. The dynamical horizon is a spacelike hypersurface  foliated by marginally trapped regions. Using this definition, they prove an important result stating that the area of the dynamical horizon always increases. They also defined timelike membranes where the evolving horizon is timelike. 
Hayward in his paper \cite{hayward1994general} has refined the concept of trapping horizons based on a 2+2 decomposition framework which introduced different types of horizons like future outer trapped horizon (FOTH), future inner trapped horizon (FITH), past outer trapped horizon (POTH) and past inner trapped horizon (PITH). There are many works \cite{booth2005marginally, helou2017causal, raviteja2020aspects, raviteja2021causal} where solutions are found for dynamical horizons, timelike membranes, and situations where an evolving horizon makes a transition from a dynamical horizon to a timelike membrane were discussed. Bousso in \cite{bousso1999holography} has introduced the construction of past holographic screens which are to be defined in terms of marginally trapped surfaces, and in works with Engelhardt \cite{bousso2015new, bousso2015proof} they have proved a new area law in general relativity where the area of holographic screens follow a  monotonic evolution even though the causal nature of these screens changes during its dynamics.
\\

The thermodynamics of evolving horizons is a work in progress, as the evolution of these horizons is a non-equilibrium phenomenon. An evolving horizon can be a dynamical horizon or a timelike tube depending on its causal nature and dynamical horizons tend to increase in their area while time-like tubes tend to decrease in their area. In general, the dynamical horizons are outer horizons (FOTH) while the timelike tubes are inner (FITH). Dynamical horizons are more generic while timelike tubes occur in special circumstances like Friedmann-Robertson-Walker (FRW) spacetime. There are therefore fundamental differences between the nature of dynamical horizons and timelike tubes. It is reasonable to expect that some thermodynamic properties, too, would carry over the distinction between dynamical horizons and timelike membranes. There have been various definitions for the surface gravity for the dynamically evolving marginally trapped regions. We think that a good formulation of surface gravity is crucial when one wants to define non-equilibrium thermodynamic state variables for various astrophysically realistic cases of evolving horizons. A few of the proposals for surface gravity that are well known are, Kodama-Hayward surface gravity \cite{hayward2004energy}, Hayward's trapping horizon \cite{hayward1994general}, Fodor et al. surface gravity \cite{Fodor}, Booth and Fairhurst surface gravity for the evolving horizon \cite{boothandfairhurst}, these are well described in \cite{Nielsen_2008}. The surface gravity expressions of various proposals for the case of the general spherically symmetric metric are in Eddington-Finkelstein coordinates and are written in terms of Misner-Sharp mass, metric function, and their derivatives with respect to these coordinates. In the paper \cite{Nielsen-2011}, they discuss a few of the proposals for surface gravity and obtain expressions for surface gravity using Painleve-Gullstrand coordinates and highlight the differences between the proposals in a dynamical setting.
\\

It is found in various solutions that the evolving horizon may transition from being spacelike to timelike, and vice versa \cite{booth2005marginally,helou2017causal, raviteja2020aspects,raviteja2021causal}. When one wants to study the thermodynamic aspects of the evolving horizons, it would be useful to understand the behavior of various surface gravity proposals with respect to the nature of causal transitions of the evolving horizons. The first paper that addresses the issue of surface gravity and causal description of evolving horizons is by \cite{faraoni-prd2011}. The causal nature of an evolving horizon in Friedmann-Lemaitre-Robertson-Walker (FLRW) spacetime is well known. For the case of FLRW, the paper \cite{faraoni-prd2011}, evaluate the Kodama-Hawyard surface gravity and shows that the surface  gravity is sensitive to the causal nature of the evolving horizon. We want to address this question in more a general context since FLRW is a specific case restricted to the cosmological type solutions.
\\

Our goal in this article is twofold. Firstly we consider the definitions of various surface gravity proposals in the D-dimension. We show that for the case of $D$ dimensional evolving marginally trapped surface, it is possible to simplify and obtain elementary formulae for the surface gravity ideas. We describe these formulae in simple terms of area radius $R$, the cosmological constant $\Lambda$, the dimension D and the principal values of the energy-momentum tensor. These formulae indicate that the surface gravity estimation can be done only using local information at the evolving horizon and does not depend upon any non-local information and the global aspects of the solution. These formulae are obtained directly from simplifying the expressions using Einstein's equations and do not require the solutions of these Einstein's equations to define the surface gravity. We obtain these formulae for the proposals of surface gravity by Kodama-Hayward \cite{hayward2004energy}, Fodor et al. \cite{Fodor}, Booth-Fairhurst \cite{boothandfairhurst} and also obtain expressions for trapping gravity by Hawyard \cite{hayward1994general}. Secondly, we want to find out which definitions of surface gravity are causally sensitive to the transitions among spacelike and timelike surfaces. Using the general formula for each proposal, we find the relation between the expression for the causal nature of the evolving horizon with the signature for surface gravity. We find that Kodama-Hayward surface gravity gives a positive value for the case of dynamical horizons and gives a negative value if the evolving horizon is timelike.
\\

\section{Causal nature of evolving marginally trapped surface}
In this section, we review some of the results describing the causal aspects of marginally trapped regions. The results can be found in \cite{faraoni-prd2011, helou2017causal,raviteja2020aspects, faraoni2012correspondence, booth2005marginally}. Some of these results generalized to a $D$ dimensional scenario for the case of spherically symmetric perfect fluid are derived in \cite{raviteja2021causal}. In the above references, the criteria for the marginally trapped surface to be timelike/null/spacelike is derived. Interestingly the causal nature of the marginally trapped surface can be obtained using Einstein equations without explicitly solving for the metric. 

We assume a general metric for a $(D = n+2)$ dimensional spherically symmetric spacetime is of the form
    \begin{equation}\label{metric}
    ds^2 = - e^{\sigma(t,r)}dt^2 + e^{\lambda(t,r)}dr^2 + R^2 (t,r)~d{\Omega}_{n}^{2}
    \end{equation}
where $d{\Omega}_{n}^{2}$ is the metric on an $n$ dimensional sphere of unit radius with angular coordinates defined by $(\theta_1, \theta_2, ..., \theta_n)$. Here, $t$ is the time coordinate, r is the comoving radial coordinate, and $R(t,r)$ is the areal radius (we will also refer to this as ``physical radius") of the n-dimensional sphere. The advantage of comoving coordinates is that the metric remains regular across the apparent horizon and becomes singular when the curvature singularity forms (this statement is generally true as long as the initial conditions are such that there are no shell crossing singularities). The matter we consider here is a perfect fluid whose energy-momentum tensor is
    \begin{equation}
    T_{\mu\nu} = (\rho(t,r) + p(t,r))u_{\mu}u_{\nu} + p(t,r) g_{\mu\nu}   
    \end{equation}
and the four-velocity in comoving coordinates is
    \begin{equation}
    u^{\mu} = (e^{-\frac{\sigma}{2}},0,0,...,0)    
    \end{equation}
with $u^{\mu} u_{\mu} = -1$. 
The relevant Einstein equations and the relations obtained by the conservation of energy-momentum tensor are given in Appendix A. It is easily seen that the energy-momentum tensor of the perfect fluid is diagonal in this coordinate system. As shown in \cite{raviteja2021causal}, the formula for the causal nature of the marginally trapped region can be expressed completely in terms of coordinate invariants and the principal values of the energy-momentum tensor ($\rho$ and $p$ in this context). 
To define the marginally trapped regions, 
for the assumed metric (\ref{metric}), we define the future outgoing radial null vector as
    \begin{equation}
    k^a = (e^{-\frac{\sigma}{2}},e^{-\frac{\lambda}{2}},0,0....,0)
    \label{outgoing}
    \end{equation}
and the future incoming radial null vector as
    \begin{equation}
    l^a = (e^{-\frac{\sigma}{2}},-e^{-\frac{\lambda}{2}},0,0....,0)
    \label{incoming}
    \end{equation}
These null vectors are normalized as 
    \begin{equation*}
    g_{ab}k^{a}l^{b} = -2
    \end{equation*}
Using these two null vectors, the induced metric on a codimension $D-2$ hypersurface that is orthogonal to the two null vectors is given by,  
    \begin{equation*}
    h_{ab} = g_{ab} + \frac{1}{2}(k_{a}l_{b} + l_{a}k_{b})
    \end{equation*}
so the expansion for the congruence of outgoing null rays is 
    \begin{equation}
    \Theta_{k} =  h^{ab} \nabla_{a} k_{b} = \frac{n}{R} \bigg( e^{-(\frac{\sigma}{2})} \dot{R} + e^{-(\frac{\lambda}{2})} R' \bigg) 
    \label{thetak}
    \end{equation}
and for completeness, the expansion for the congruence of incoming null rays is 
    \begin{equation}
    \Theta_{l} =  h^{ab} \nabla_{a} l_{b} = \frac{n}{R} \bigg( e^{-(\frac{\sigma}{2})} \dot{R} - e^{-(\frac{\lambda}{2})} R' \bigg) 
    \end{equation}
The hypersurface given by the equation $\Theta_{k} = c$ is a curve foliated by a marginally trapped region if we set the constant $c = 0$, so the curve is a marginally trapped tube. To obtain the causal nature of the curve, we find the norm of normal $\beta_k$ to this curve evaluated at $\Theta_k=0$ and is given by \cite{raviteja2021causal},
    \begin{equation}
    \beta_{k} = - \bigg( (\pounds_{k} \Theta_{k}) (\pounds_{l} \Theta_{k}) \bigg) \bigg|_{\Theta_{k} = 0} 
    \label{productoflie}
    \end{equation}
The lie derivative of $\Theta_{k}$ with respect to the outgoing null vector $k^{a}$ and the incoming null vector $l^{a}$ are
    \begin{equation} \label{explkok}
    \pounds_{k} \Theta_{k} = k^{a} \nabla_{a} \Theta_{k} = e^{\frac{-\sigma}{2}} \partial_{t} \Theta_{k} + e^{\frac{-\lambda}{2}} \partial_{r} \Theta_{k}
    \end{equation}
    \begin{equation} \label{expllok}
    \pounds_{l} \Theta_{k} = l^{a} \nabla_{a} \Theta_{k} =  e^{\frac{-\sigma}{2}} \partial_{t} \Theta_{k} - e^{\frac{-\lambda}{2}} \partial_{r} \Theta_{k}
    \end{equation}
These lie derivatives have to be evaluated at $\Theta_{k} = 0$ 
    \begin{equation}
    \pounds_{k} \Theta_{k} \bigg|_{\Theta_{k} = 0} = - \tilde{\kappa}  (\rho + p)
    \end{equation}
and
    \begin{equation}
    \pounds_{l} \Theta_{k} \bigg|_{\Theta_{k} = 0} = \tilde{\kappa} (\rho - p) + 2\Lambda - \frac{n (n-1)}{R^2}
    \label{liederivativeone}
    \end{equation}
which gives us
    \begin{equation} \label{betak}
    \beta_{k} = \tilde{\kappa}  (\rho + p) \bigg( \tilde{\kappa} (\rho - p) + 2\Lambda - \frac{n (n-1)}{R^2} \bigg)
    \end{equation}

Instead of the normal, one can also obtain the causal nature of the curve $\Theta_k=0$  using the ratio of lie derivatives. The ratio  represents the causal nature of the tangent to the curves that are foliated by marginally trapped region, the proof of which is shown in \cite{hayward1994general,dreyer2003introduction}. The ratio of the Lie derivatives evaluated at $\Theta_{k} = 0$ determines the causal nature of the marginally trapped tube, which is  
    \begin{equation} \label{alphak}
    \alpha_{k} = \frac{\pounds_{k} \Theta_{k}}{\pounds_{l} \Theta_{k}} \bigg|_{\Theta_{k} = 0} = \frac{- \tilde{\kappa}  (\rho + p)}{\tilde{\kappa}  (\rho - p) + 2\Lambda - \frac{n (n-1)}{R^2}}
    \end{equation}
The causal nature is described by the sign of the norm of the normal or tangent. Marginally trapped curve is timelike if $\beta_k>0$ , is spacelike if $\beta_k<0$ and is null if $\beta_k=0$. We can see that $\beta_{k}$ and $\alpha_{k}$ are always of opposite signs. So the causal criteria are reversed for $\alpha_k$.
\\

We note the fact that the formula for $\beta_k$ is completely local and does not need the solution of the Einstein equations. The formula is described in terms of geometric invariants and system parameters like area radius $R$ of the marginally trapped surface, cosmological constant $\Lambda$, dimension of the spacetime ($n=D-2)$ and the principal values of the energy-momentum tensor $(\rho,p)$ at the location of the marginally trapped region. One can easily see that by adjusting the density and pressure, one can obtain transitions of the marginally trapped tube from timelike to spacelike and vice versa. We also note that the norm of the normal is a better tool to look at these causal transitions since $\beta_k$ goes to zero and hence regular while $\alpha_k$ goes to infinity and therefore is analytically cumbersome at the transition points.
\\

Now we will analytically describe these transitions in the FRW spacetime setting. 
\\
{\it{FRW case}}: The metric (\ref{metric}) can be brought to the standard FRW form,
    \begin{equation} \label{ndimhommetric}
    ds^2 = - dt^2 + a^{2}(t) \bigg(\frac{dr^2}{1-kr^2}  + r^2~d{\Omega}_{n}^{2} \bigg)
    \end{equation}
which is the higher dimensional spherically symmetric metric whose source is a homogeneous perfect fluid. The function $a(t)$ has the standard interpretation as the scale factor and $k$ takes the values in  $(1,0,-1)$.
For the metric (\ref{ndimhommetric}), the future incoming radial null vector is given by
    \begin{equation}
    l^a = (1,-\frac{\sqrt{1 - k r^2}}{a(t)},0,0....,0)
    \end{equation}
and the future outgoing radial null vector is given by
    \begin{equation}
    k^a = (1,\frac{\sqrt{1 - k r^2}}{a(t)},0,0....,0).
    \end{equation}
These are normalized to, 
    \begin{equation*}
    g_{ab}k^{a}l^{b} = -2.
    \end{equation*}
The expansion scalar for the outgoing bundle of null rays is 
    \begin{equation} \label{homthetak}
    \Theta_{k} =  h^{ab} \nabla_{a} k_{b} = \frac{n }{a(t)r} \bigg( \dot{a}(t) r + \sqrt{1 - k r^2} \bigg) 
    \end{equation}
    
We know that the norm of the normal to the $\Theta_{k} = 0$ curve can be expressed as product of Lie derivatives (\ref{productoflie}). The lie derivative of $\Theta_{k}$ with respect to the outgoing radial null vector is
    \begin{equation}
    \pounds_{k} \Theta_{k} \bigg|_{\Theta_{k}=0} = -\tilde{\kappa}  \rho ( 1 + \omega)
    \end{equation}
and, that with respect to the ingoing radial null vector is
    \begin{equation}
    \pounds_{l} \Theta_{k} \bigg|_{\Theta_{k}=0} = \frac{n}{2R^2}\left(3-n-\omega(n+1)+\frac{2\Lambda R^2 (1+\omega)}{n}\right)
    \label{liederivativetwo}
    \end{equation}
The norm of the normal to the curves $\Theta_{k}=0$ can  be expressed as
    \begin{equation}   \label{hombeta}
    \beta_{k} = \frac{n \tilde{\kappa} \rho ( 1 + \omega)}{2R^2}
    \left( 3-n-\omega(n+1)+\frac{2\Lambda R^2 (1+\omega)}{n} \right) 
    \end{equation}
As described earlier, the causal nature of the marginally trapped tube can also be found using the ratio of lie derivatives, which represents the causal nature of the tangent of the marginally trapped tube. The ratio of the lie derivatives evaluated at $\Theta_{k} = 0$ and $\Theta_{l} = 0$ gives us
    \begin{equation} \label{homalpha}
    \alpha_{k} = \frac{-2 R^2 \tilde{\kappa}  \rho ( 1 + \omega)}{n\left(3-n-\omega(n+1)+\frac{2\Lambda R^2 (1+\omega)}{n}\right)}
    \end{equation}
Note that we have expressed the formula in terms of physical radius $R$ instead of $a(t)r$. The formula for the cosmological case is even more elementary than the general case. 
If we assume $1+\omega$ is positive, the sign of the expressions for $\beta_k$ and $\alpha_k$ are completely determined by the following expression,
    \begin{equation} \label{criticalterm}
    \left( 3-n-\omega(n+1)+\frac{2\Lambda R^2 (1+\omega)}{n} \right)
    \end{equation}
We note that the formula does not contain any dynamical variables of the model and is expressed completely in terms of spacetime dimension ($n$), equation of the state parameter ($\omega$), cosmological constant ($\Lambda$), and physical radius ($R$) in this sense it is a geometrical result.
\\

If we consider the case where the cosmological constant is zero, we see from the above expression that as the marginally trapped region evolves, there is no change of causal nature. It is uniformly timelike, spacelike, or null. At a value of $\omega$ called $\omega_{critical}$ given below, the evolving marginally trapped region is uniformly null. If $\omega<\omega_{critical}$, it is timeline, and if $\omega>\omega_{critical}$, the marginally trapped tube is spacelike.
    \begin{equation}
    \omega_{critical} = - \frac{(n-3)}{(n+1)}
    \end{equation}
The case where the equation of state coincides with $\omega_{critical}$ is very special. The evolving marginally trapped tube is uniformly null. These are called the null evolving horizons that do not fit in Hayward's classification criteria as shown in \cite{raviteja2021causal}.

If we consider the case, $\Lambda \neq 0$  there exists a critical radius $R_{critical}$ at which marginally trapped tube is null ($\beta = 0 = \alpha^{-1}$) and it also marks a transition of these curves from spacelike region to timelike region or vice versa. 
    \begin{equation*}
    R_{critical}^{2} = \frac{n(n-3) + n\omega (n+1)}{2 \Lambda (1+\omega)}
    \label{rcritical}
    \end{equation*}
Whenever the marginally trapped tube crosses this critical radius, it makes a transition in terms of its causal nature.

\section{Surface gravity of marginally trapped surfaces}
In this section, we study the various proposals that define surface gravity in a dynamic setting. For almost all the proposals, we obtain elementary formulae where the surface gravity is obtained in terms of invariants and local information like area radius, the cosmological constant, the number of dimensions, and principal values of the energy-momentum tensor. These simplified formulae are intuitively appealing and have not been reported before in literature. With the help of these formulae, it is easy to compare the behavior of various proposals for surface gravity for marginally trapped regions with the causal behavior of the same. We note a useful result that the surface gravity of $D$ dimensional Schwarzschild black hole is, $\kappa =(D-3)/2R=(n-1)/2R$. This helps in fixing the arbitrary constant that is due to the freedom of normalization of null rays. 

\subsection{Kodama-Hayward Surface Gravity}
The paper \cite{faraoni-prd2011}, has worked on Kodama-Hayward surface gravity for the case of marginally trapped surfaces in FRW spacetimes for the case of $D=4$ dimensions. The paper finds that Kodama-Hayward's definition of surface gravity is sensitive to the causal description of the evolving marginally trapped surface. It is shown in the paper that for the FRW case, if we define the perfect fluid with the equation of state given by, $\rho = \omega P$, for $\omega<1/3$, the marginally trapped surface is timelike. The Kodama-Hayward surface gravity is shown in \cite{faraoni-prd2011} to be negative in this range. For $\omega>1/3$, the marginally trapped surface is spacelike, and Kodama-Hayward surface gravity is shown to be positive. The expressions for Kodama-Hayward surface gravity for the dynamical scenario using Painleve-Gullstrand coordinates were obtained in \cite{Nielsen-2011} in terms of derivatives of the Schwarzschild mass.  We now obtain a formula for a spherically symmetric scenario in $D$ dimensions for a perfect fluid. The Kodama vector for the spherically symmetric spacetime generalized to $D$ dimensions is defined as,
    \begin{equation}
    K^{\mu}=\frac{1}{\sqrt{-h}}\epsilon^{\mu\nu}\partial_{\nu}R
    \end{equation}
where $R$ is the areal radius and $-h$ is the determinant of the metric induced on the horizon $h_{ab}$ (on the two dimensional hypersurface orthogonal to the $D-2$ dimensional sphere).

    \begin{equation}
    \begin{split}
    \kappa_{K-H} = \frac{C_1}{\sqrt{-h}}\epsilon^{\alpha}_{\mu}\partial_{\alpha}K^{\mu}=\\
    \frac{C_1}{\sqrt{-h}} \frac{\partial}{\partial x^\mu}  \bigg(\sqrt{-h} h^{\mu\nu}  \frac{\partial}{\partial x^{\nu}} R \bigg)
    \end{split}
    \end{equation}

The constant $C_1$ is dependent on the normalization of the Kodama vector and is fixed indirectly by matching the value of the surface gravity for the known static case in D dimensions.  For the metric (\ref{metric}), using 
$\sqrt{-h}$ = e$^{\sigma/2}$ e$^{\lambda/2}$ and evaluating the above expression, we obtain,

\begin{equation}
  \kappa_{K-H} = C_1 \bigg(-e^{-\sigma} ( \ddot{R} + \dot{R}( \frac{\dot{\lambda} -\dot{\sigma} }{2}) ) + e^{-\lambda} ( R'' + R'\frac{(\sigma'-\lambda')}{2} ) \bigg) 
  \label{usefuleqn3}
\end{equation}

where, $\dot{R} = \frac{\partial R}{\partial t} $ and $R'= \frac{\partial R}{\partial r}$.
We show that the complicated expression can, surprisingly, be simplified to a simple form using Einstein equations for the marginally trapped regions. 

For the apparent horizon, the outgoing null vector is given by (\ref{outgoing}). The condition for the outgoing null ray to be marginally trapped is obtained by setting $\Theta_k$ to zero. 
    \begin{equation}
    \Theta_{k} = \frac{n}{R} \bigg( e^{-(\frac{\sigma}{2})} \dot{R} + e^{-(\frac{\lambda}{2})} R' \bigg) = 0
    \end{equation}
This gives 
    \begin{equation}
    \dot{R} e^{-\sigma/2} = -R'e^{-\lambda/2}
    \label{usefulterm}
    \end{equation}
Using the above relation, we can write
    \begin{equation}
    -e^{-\sigma} \frac{\dot{R} \dot{\lambda}}{2} = e^{-\frac{(\sigma + \lambda)}{2}} \frac{R' \dot{\lambda}}{2}
    \label{usefulterm1}
    \end{equation}
and
    \begin{equation}
    e^{-\lambda} \frac{R' \sigma'}{2} = -e^{-\frac{(\sigma + \lambda)}{2}} \frac{\dot{R} \sigma'}{2}
    \label{usefulterm2}
    \end{equation}
We now simplify equation \ref{usefuleqn3} using the Einstein equations given in Appendix A. We calculate $G_{00}-G_{11}$ (using equations \ref{G00} and \ref{G11} to obtain the expression \ref{G00-G11}. We simplify the equation  \ref{G00-G11}, making use of the results \ref{usefulterm},\ref{usefulterm1}, \ref{usefulterm2}). The expression \ref{usefuleqn3} simplifies to the equation below. 
    \begin{equation}
    \kappa_{K-H} = C_{1} \bigg( \frac{n-1}{R} - \frac{R}{n}(\tilde{\kappa}(\rho - p) + 2 \Lambda )  \bigg)
    \label{kodamahayward}
    \end{equation}
Comparing the result for zero density, pressure, and the cosmological constant, with the surface gravity of D-dimensional Scwarzschild spacetime, we get $C_1=1/2$. We observe that the Kodama-Hayward surface gravity is completely determined by the local information available at the marginally trapped region. We note that if the apparent horizon is isolated, then the curve $\Theta_k=0$ is null. We set $(\rho=0, p=0)$ in the above equation and see that the surface gravity in $D$ dimensions is proportional to $1/R$. We recover the surface gravity of the static black hole horizon when the formula is adopted for the non-evolving scenario. We also recover the surface gravity of the de Sitter horizon or cosmological horizon by setting $\rho=0, p=0$. We see that the surface gravity of the de Sitter cosmological horizon in a pure de Sitter spacetime is $-\sqrt{2\Lambda}/\sqrt{n(n+1)}$. 
For FRW case, using reference \cite{PhysRevD.103.124005}, we obtain the following formula for the perfect fluid for the equation of state given below, 
    \begin{equation} 
    p = \omega \rho.
    \end{equation}
    \begin{equation}
    \kappa_{K-H} = \frac{1}{4} \bigg ( \frac{n-3+\omega (n+1)}{R} - \frac{2 \Lambda R}{n} (1+\omega) \bigg)
    \label{kodamahaywardhomogeneous}
    \end{equation}
We note that this formula is derived by simplifying the expressions for the cosmological case. Setting $\omega=0$ gives us the formula for dust. This formula is valid for FRW spacetime ($\rho\neq0$).

\subsubsection{Causal correlation of Kodama-Hayward surface gravity}
In the previous section, we observed the possibilities of transitions of an evolving horizon from spacelike to timelike and vice versa. We now examine the value of surface gravity as the marginally trapped region changes from timelike to spacelike. To find this relation, we divide the surface gravity in \ref{kodamahayward} with \ref{betak}. We get,
    \begin{equation}
    \frac{\kappa_{K-H}}{\beta_k}=\frac{  \bigg( \frac{n-1}{R} - \frac{R}{n}(2\tilde{\kappa}(\rho - p) + 2 \Lambda )  \bigg)}{\tilde{\kappa} (\rho + p) \bigg( \tilde{\kappa} (\rho - p) + 2\Lambda - \frac{n (n-1)}{R^2} \bigg)}
    \end{equation}
The ratio can be simplified to be $- R/(2n\tilde{\kappa}(\rho+p))$. If we assume the energy condition that $\rho+p>0$, we see that the ratio is a negative definite quantity. This implies that always the surface gravity is negatively correlated with the norm of the normal to the marginally trapped region. This implies that for the dynamical horizon where the marginally trapped curve is spacelike, the Kodama-Hayward surface gravity is also positive. For the case when the marginally trapped curve is timelike that is for timelike tubes, the surface gravity is negative.
\\

In the FRW case, an interesting class of solutions where the evolving marginally trapped region is null. These are degenerate cases that escape the classification criteria of evolving marginally trapped regions into outer and inner horizons \cite{raviteja2021causal}.  For four dimensions, this corresponds to an equation of state given by $p=\rho/3$ and for a general dimension $D=n+2$, we have $p=- \rho(n-3)/(n+1)$. The surface gravity corresponding to these cases is $\kappa_{K-H}=0$.

Also, in the presence of a cosmological horizon, the evolving marginally trapped region transitions from spacelike to timelike when the curve passes through the critical radius given by (\ref{rcritical}). So the surface gravity too transitions from a positive value to a negative value passing through zero. The nature of transitions needs to be explored further and is left for future consideration.

\subsection{Hayward's trapping gravity}
Another useful quantity defined by Hayward \cite{hayward1994general}, is called trapping gravity. It is defined below.
    \begin{equation}
    \kappa_H = \frac{1}{2} \sqrt{-l^\alpha \Theta_{k;\alpha}} = \frac{1}{2} \sqrt{-\pounds_l \Theta_k}
    \end{equation}

We now provide a formula for estimating trapping gravity using equation \ref{liederivativeone}, (from \cite{PhysRevD.103.124005})
\begin{equation}
\kappa_H = \frac{1}{2} \sqrt{\frac{n(n-1)}{R^2} -2\Lambda - K (\rho - p)}
\label{haywardtrapping}
\end{equation}
for a general spherically symmetric perfect fluid scenario. 

For the FRW  case, we obtain the formula below for a perfect fluid with the equation of state given by, $p=\omega \rho$. Using (\ref{liederivativetwo}), we obtain the formula for trapping horizon in FRW gravity to be, 
\begin{equation}
\kappa_H = \frac{1}{2} \sqrt{\frac{\bigg(n-3+\omega(n+1)\bigg) n}{2R^2} - \Lambda(1+\omega)}
\label{haywardtrappinghomogeneous}
\end{equation}
As can be seen from both the expressions of Hayward gravity, these are defined only for the case of dynamical horizons and not timelike tubes. This is because of the square root. The expression inside the square root is positive (assuming the null energy condition is preserved) only when the marginally trapped region is a spacelike curve and is negative if the marginally trapped region is described by a timelike curve. We note that the expressions inside the square root are causally sensitive in the same sense as the Kodama-Hayward surface gravity. The value of Hayward's trapping horizon becomes imaginary for timelike tubes.

\subsection{Fodor's method}
We now consider the definition of surface gravity due to Fodor et al. \cite{Fodor}. We again try to obtain a formula for Fodor's surface gravity. Obtaining an expression for the surface gravity of a perfect fluid is found to be analytically difficult. We instead obtain a formula where the marginally trapped region is evolving due to dust. We, therefore, set the pressure to zero. We find that the expression simplifies to give a closed form. We then use the obtained expression to find the causal correlation. We recall the definition of Fodor's surface gravity as, 
    \begin{equation}
    \kappa_F = -l^\beta k^\alpha k_{\beta;\alpha}
    \end{equation}
with the same definitions of $k$ and $l$ as given in the earlier sections. The normalization prescribed in Fodor et al. \cite{Fodor}, is given by
    \begin{equation*}
    l^\alpha k_\alpha = -1
    \end{equation*}
However, we work with a different normalization and finally decide on the normalization by matching the surface gravity that is defined up to a multiplicative constant, with the static black hole case. For the metric defined in (\ref{metric}), we get the surface gravity to be,
    \begin{equation}\label{kf}
    \kappa_F = \frac{C_2}{2 \sqrt{2}} \bigg(e^{-\lambda/2} \sigma' + e^{-\sigma/2} \dot{\lambda} \bigg)
    \end{equation}
where $C_2$ is fixed by comparing with the known static scenario. For the case of dust, the metric \ref{metric} becomes that of a D-dimensional  Lemaitre Toman Bondi (LTB) model. Comparing with the D-dimensional LTB model, \cite{raviteja2020aspects}, we get the metric coefficient $\sigma = 0 $ and, 
    \begin{equation}\label{elambda}
    e^\lambda = R'^2
    \end{equation}
We simplify the expression (\ref{kf}) using the above relation. We get,
    \begin{equation}
    \kappa_F = \frac{C_2}{2 \sqrt{2}} \dot{\lambda}
    \end{equation}
Now, from the expression(\ref{elambda}) we have
    \begin{equation*}
    \dot{\lambda} = \frac{2 \dot{R}'}{R'}   
    \end{equation*}
Using the standard relation for the LTB models \cite{raviteja2020aspects}
    \begin{equation}\label{rdotsquare}
    \dot{R}^2 = \frac{F(r)}{R^{n-1}} + \frac{2 \Lambda R^2}{n(n+1)}  
    \end{equation}
where $F(r)$ is a function only of the comoving radius $r$ and has the interpretation of Misner-Sharp mass. It is the total mass within a shell with comoving label $r$.
For marginally trapped and anti-trapped surfaces, it is shown in \cite{raviteja2020aspects} that $\dot{R}^2 = +1$ holds true. Here $\dot{R}$ is the derivative of the area radius with comoving time. For the marginally trapped case, we choose, $\dot{R} = -1$. We also have from reference \cite{raviteja2020aspects},
    \begin{equation}
    \frac{F'}{2 R^n R'} = \frac{\tilde{\kappa} \rho}{n}
    \end{equation}
Now by differentiating  \ref{rdotsquare} w.r.t $r$, we arrive at the expression
    \begin{equation}
    2 \dot{R} \dot{R}' = \frac{F'}{R^{(n-1)}} + \frac{4 \Lambda R R'}{n(n+1)} - (n-1) \frac{F R'}{R^n}
    \end{equation}
Dividing both sides by $R'$, we arrive at,
    \begin{equation}\label{rdotprimebyrprime}
    \frac{\dot{R}'}{R'} = R \bigg( \frac{- \tilde{\kappa} \rho}{n} - \frac{2 \Lambda}{n(n+1)} + \frac{(n-1)F}{2 R^{(n+1)}} \bigg)
    \end{equation}
For the marginally trapped region, we set $\Theta_k=0$. This implies \cite{raviteja2020aspects}, 
    \begin{equation}
    \frac{F}{R^{n+1}} = \frac{1}{R^2} -\frac{2 \Lambda}{n(n+1)} 
    \end{equation}
Using the above expression, equation (\ref{rdotprimebyrprime}) gives 
    \begin{equation}
    \frac{\dot{R}'}{R'} = \frac{n-1}{2R} - R \bigg( \frac{\tilde{\kappa} \rho}{n} + \frac{\Lambda}{n} \bigg)
    \end{equation}
Using this, we arrive at the expression for surface gravity to be, 
    \begin{equation}
    K_{F} = C_2 \bigg( \frac{n-1}{2R} - \frac{R}{n} ( \tilde{\kappa} \rho + \Lambda)\bigg)
    \label{kfodor}
    \end{equation}
Compared with $D$ dimensional Schwarzschild case, we get that $C_2=1$. We see that the equation (\ref{kfodor}) matches with the result in \cite{Fodor} if we set the dimension $D=4(n=2)$ and equate the cosmological constant to zero. In this sense, the result obtained is a generalization of the result in \cite{Fodor} to $D$ dimensions and with a cosmological constant term.

\subsubsection{Causal correlation of $\kappa_F$}
For correlating with the causal nature of the marginally trapped surface, we compare the expression for the surface gravity with the formula for causal nature. Since the formula obtained in the Fodor et al. case is only for dust, we adopt the formula by setting the pressure to zero.
    \begin{equation}
    \frac{\kappa_F}{\beta_k}=\frac{\bigg( \frac{n-1}{2R} - \frac{R}{n} ( \tilde{\kappa} \rho + \Lambda)\bigg)}{\tilde{\kappa} \rho \bigg( \tilde{\kappa} \rho + 2\Lambda - \frac{n (n-1)}{R^2} \bigg)}
    \end{equation}
We can rearrange the denominator to arrive at the,
    \begin{equation}
    \frac{\kappa_F}{\beta_k}=\frac{-R\bigg( \frac{n-1}{2R} - \frac{R}{n} ( \tilde{\kappa} \rho + \Lambda)\bigg)}{2n\tilde{\kappa} \rho  \bigg( \frac{n-1}{2R} - \frac{R}{n} (\frac{\tilde{\kappa} \rho}{2} + \Lambda)\bigg)}
    \end{equation}
Again assuming that $\rho+p>0$, we observe that there is a factor of 2 as the coefficient in the density term $\rho$ that makes a difference from $\kappa_F$ being causally correlated. The ratio $\frac{\kappa_F}{\beta_k}$ is therefore not negative definite and hence the relation between the causal nature of marginally trapped region breaks down. However, there is a match between $\kappa_{KH}$ and $\kappa_F$ when the energy density is zero. Also, there is an agreement between $\kappa_{KH}$ and $\kappa_F$ when the energy densities are small or when the area radius of the marginally trapped region is small.

\subsection{Booth and Fairhurst method}
The dynamical surface gravity is defined in the paper \cite{boothandfairhurst}. This definition is tailored for the case of a slowly evolving horizon. The criteria for the same is described in \cite{boothandfairhurst}. Using the idea in \cite{boothandfairhurst}, we can define the tangent to the marginally trapped curve to be $V^{\alpha}=k^\alpha-\alpha_k l^{\alpha}$ and normal to be $\tau_{\alpha}=k_{\alpha}+\alpha_k l_{\alpha}$. The norm for each vector is evaluated with a normalization $k^{\alpha}l_{\alpha}=-2$ is given by $\alpha_k$ (\ref{alphak}) and $\beta_k$ (\ref{betak}) respectively. We define surface gravity to be,
    \begin{equation}\label{kbf}
    K_{BF} = - l^\alpha k^\beta k_{\alpha;\beta} - \alpha_k k^\alpha l^\beta l_{\alpha;\beta}
    \end{equation}
To evaluate the above expression, we have
    \begin{equation*}
    l^\beta k^\alpha k_{\beta;\alpha} = -C_3 \bigg(e^{-\lambda/2} \sigma' + e^{-\sigma/2} \dot{\lambda} \bigg)
    \end{equation*}
    \begin{equation*}
    k^\alpha l^\beta l_{\alpha;\beta} = C_3 \bigg(e^{-\lambda/2} \sigma' - e^{-\sigma/2} \dot{\lambda} \bigg)
    \end{equation*}
after putting the metric coefficient $\sigma$ = 0, we get 
    \begin{equation}
    l^\beta k^\alpha k_{\beta;\alpha} = -C_3 \dot{\lambda}
    \end{equation}
and
    \begin{equation}
    k^\alpha l^\beta l_{\alpha;\beta}   = -C_3 \dot{\lambda}
    \end{equation}
Hence equation (\ref{kbf}) gives us
    \begin{equation}
    K_{BF} = C_3 \bigg(1 + \alpha_k\bigg) \dot{\lambda}
    \end{equation}
Just as in the previous section, we evaluate the surface gravity for the analytically tractable case of dust. We set the pressure to zero. We then obtain the equation for the scalar $\alpha_k$ to be
    \begin{equation}
    \alpha_k = \frac{- \tilde{\kappa} \rho}{\tilde{\kappa} \rho + 2 \Lambda - \frac{n(n-1)}{R^2}}
    \end{equation}
and $\dot{\lambda}$ is evaluated in Fodor's method. This yields  
    \begin{equation}
    \kappa_{BF} = \frac{C_3 \bigg( \frac{n-1}{2R} - \frac{R}{n} ( \tilde{\kappa} \rho + \Lambda )\bigg) \bigg( 2 \Lambda - \frac{n(n-1)}{R^2}\bigg)}{\tilde{\kappa} \rho + 2 \Lambda - \frac{n(n-1)}{R^2}}
    \label{booth}
    \end{equation}
We see again that the expressions match with the static case of $D$ dimensional Schwarzschild black hole surface gravity where we set density $\rho=0$ and $\Lambda=0$. Comparison sets the value of $C_3=1$. The result matches with D-dimensional Schwarzschild deSitter surface gravity.

\subsubsection{Causal correlation of the Booth-Fairhurst surface gravity}
An inspection of (\ref{booth}) indicates that the denominator of the expression is positive definite with respect to $\beta_k$, but the numerator is not. This makes the expression $\kappa_{BF}$ not causally correlated. However, for the slowly evolving scenario for which the expression was the originally intended purpose of defining the surface gravity \cite{boothandfairhurst}.

\section{Conclusions}
In this article, we consider the marginally trapped surfaces which can be spacelike, timelike, or null depending on the local information at the horizon, i.e density $\rho$, pressure p, area radius $R$, and cosmological constant $\Lambda$. We analyze the various ideas regarding the definition of surface gravity when the horizon is evolving. We derived the following simple formulae for the various proposals. The Kodama-Hayward
surface gravity for the general case is (\ref{kodamahayward}). For the case of FLRW, we obtain the formula (\ref{kodamahaywardhomogeneous}). We showed that both formulae are sensitive to causal nature. The formula yields that surface gravity is positive for dynamical horizons and negative for timelike tubes. The surface gravity smoothly transitions whenever the evolving horizon makes a transition from timelike to spacelike or vice versa. The formula obtained for Kodama-Hayward surface gravity holds  for any spherically symmetric situation with and without cosmological constant for any density and pressure of the fluid. This definition of surface gravity seems better suited to study thermodynamic aspects of evolving horizons since it is sensitive to the causal nature of the horizon. This also holds promise for future analysis between the nature of causal transitions and non-equilibrium thermodynamic state variables that could be defined on the horizons. 
\\
We obtained the formula for Hayward's trapping gravity (\ref{haywardtrapping}), and similarly, for the FLRW case the formula is derived in (\ref{haywardtrappinghomogeneous}). Based on the expressions, it is clear that this quantity is defined only for dynamical horizons and this quantity becomes imaginary for timelike tubes. We obtained the formula for Fodor et al. surface gravity in (\ref{kfodor}). This formula is obtained for the case of zero pressure and with the cosmological constant. The case with pressure is found to be analytically not tractable and a closed-form expression was not possible. Nevertheless, we could obtain the causal correlation for the Fodor et al case and find that there is a mismatch with the causal description. The parameter space where Fodor's surface gravity transitions in terms of its sign is different from the transitions in evolving horizons. We similarly obtain the formula for Booth and Fairhurst's proposal for surface gravity in (\ref{kbf}). Again the expressions were obtainable in a closed form only for the case of zero pressure. Just as in Fodor's case, the Booth and Fairhurst proposal does not correlate with the causal description of the evolving horizon in terms of changing the sign of surface gravity when one goes from spacelike to timelike evolving horizons. These findings in the article, are therefore crucial for the first steps towards defining thermodynamic variables in evolving horizons. 
\section{Appendix}
The nonzero components of the energy-momentum tensor are listed below
    \begin{eqnarray*}
    &&
    T_{00} = \rho e^{\sigma}
    \\ &&
    T_{11} = p e^{\lambda} 
    \\ &&
    T_{22} = p R^2
    \\ &&
    T_{(l+1 ~ l+1)} = sin^2{\theta_{(l-1)}} T_{(ll)} 
    \end{eqnarray*}
where $l$ takes values from 2 to n. From the Bianchi identities 
    \begin{equation}
    {T^{\mu\nu}}_{;\nu} = 0
    \end{equation}
we get the following relations
    \begin{equation} \label{timebianchi}
    \dot{\rho} + \frac{(\rho + p)}{2} \bigg(\frac{2n\dot{R}}{R} + \dot{\lambda} \bigg) = 0
    \end{equation}
    \begin{equation} \label{radialbianchi}
    p' + \frac{\sigma'}{2} (p + \rho) = 0
    \end{equation}
where the dot represents a derivative with time coordinate and the prime represents a derivative with the comoving radial coordinate.
\\

For a nonzero cosmological constant ($\Lambda \neq 0$), the Einstein equations are
    \begin{equation}
    G_{\mu\nu} + \Lambda g_{\mu\nu} = \tilde{\kappa}  T_{\mu\nu}
    \end{equation}
here $\tilde{\kappa} $ is a constant and is related to gravitational constant $G_n$, ($\tilde{\kappa}$ = 8$\pi G_n$). With these conditions, we evaluate the left-hand side components of the Einstein equation $(G_{\mu\nu} + \Lambda g_{\mu\nu})$ which are summarized below
    \begin{eqnarray}
    && G_{00}+\Lambda g_{00} = \frac{e^{-\lambda}}{R^2} \bigg[ \frac{n(n-1)}{2} (e^{\lambda + \sigma}+ e^{\lambda} \dot{R}^2 - e^{\sigma} R'^{2})  \nonumber \\ && + \frac{n}{2} R (-2 R'' e^{\sigma} + e^{\sigma} R' \lambda' + e^{\lambda}  \dot{R} \dot{\lambda}) - \Lambda e^{\lambda + \sigma} R^{2} \bigg]
    \end{eqnarray}
    \begin{equation} \label{G01}
    G_{01} +\Lambda g_{01}= \frac{n}{2}\frac{(R' \dot{\lambda} - 2 \dot{R}' +  \sigma'\dot{R} )}{R} 
    \end{equation}
    \begin{eqnarray}
    && G_{11}+\Lambda g_{11} = \frac{e^{-\sigma}}{R^{2}} \bigg[ - \frac{n(n-1)}{2} (e^{\lambda + \sigma}+ e^{\lambda} \dot{R}^2 - e^{\sigma} R'^{2})  \nonumber \\ && + \frac{n}{2} R ( e^{\sigma} R' \sigma' + e^{\lambda}(\dot{R}\dot{\sigma} - {\color{black} 2\ddot{R}})) + \Lambda e^{\lambda + \sigma} R^{2} \bigg] 
    \end{eqnarray}
    \begin{eqnarray}
    && G_{22}+\Lambda g_{22} =  \frac{e^{-(\lambda+\sigma)}}{4}  \bigg[ 2(n-1)(n-2) ( e^{\sigma} R'^{2} - e^{\lambda +\sigma}  \nonumber
    \\ &&
    - e^{\lambda}\dot{R}^2) - 2(n-1)R (e^{\sigma} R'(\lambda' - \sigma') - 2  e^{\sigma} R'' \nonumber 
    \\ && 
    + e^{\lambda}(\dot{R}(\dot{\lambda} - \dot{\sigma}) + 2\ddot{R})) + R^{2}(4e^{\lambda + \sigma} \Lambda - e^{\lambda}(2\ddot{\lambda} +\dot{\lambda}^2 \nonumber 
    \\ && 
    - \dot{\lambda}\dot{\sigma}) +  e^{\lambda}(2\sigma''+ {\sigma'}^2 - \lambda' \sigma'))  \bigg] \nonumber
    \end{eqnarray}
The other nonzero relations are given by 
    \begin{equation}
    G_{(j+1~j+1)} = sin^2{\theta_{(j-1)}} G_{(jj)}
    \end{equation}
where $j$ takes values from 2 to $n$. So from the $G_{01} = 0$ Einstein equation we get
    \begin{equation} \label{simpleG01}
    R'\dot{\lambda} - 2 \dot{R}' + \sigma'\dot{R} = 0
    \end{equation}
from the $G_{00} = \tilde{\kappa} \rho e^{\sigma}$ equation we have
    \begin{eqnarray} \label{G00}
    && 
    \frac{n(n-1)}{2} (e^{\lambda + \sigma}+ e^{\lambda} \dot{R}^2 - e^{\sigma} R'^{2})  
    \\ && 
    + \frac{n}{2} R (-2 R'' e^{\sigma} + e^{\sigma} R' \lambda' + e^{\lambda}  \dot{R} \dot{\lambda}) =  (\tilde{\kappa}  \rho +\Lambda) R^2 e^{\lambda + \sigma} \nonumber
    \end{eqnarray}
and $G_{11} = \tilde{\kappa} p e^{\lambda}$ equation we get
    \begin{eqnarray} \label{G11}
    && 
    -\frac{n(n-1)}{2} (e^{\lambda + \sigma}+ e^{\lambda} \dot{R}^2 - e^{\sigma} R'^{2})
    \\ && 
    + \frac{n}{2} R ( e^{\sigma} R' \sigma' + e^{\lambda}(\dot{R}\dot{\sigma} -  2\ddot{R}) ) =  (\tilde{\kappa} p - \Lambda) R^2 e^{\lambda + \sigma} \nonumber
    \end{eqnarray}
The following two expressions will be useful in the subsequent calculations done in the paper. The sum (\ref{G00}) + (\ref{G11}) gives,
    \begin{eqnarray} \label{G00+G11}
    && \frac{n}{2} \bigg( e^{\sigma}R'(\sigma' + \lambda') + e^{\lambda}\dot{R}(\dot{\sigma}+\dot{\lambda}) - 2 (e^{\lambda}\ddot{R} + e^{\sigma} R'')   \bigg)      \nonumber
    \\ 
    && = \tilde{\kappa} (\rho + p) e^{(\lambda + \sigma)} R
    \end{eqnarray}
similarly the difference (\ref{G00}) - (\ref{G11}) gives us
    \begin{eqnarray} \label{G00-G11}
    && 
    n(n-1) (e^{\lambda + \sigma}+ e^{\lambda} \dot{R}^2 - e^{\sigma} R'^{2})  
    \\ && 
    + \frac{nR}{2}  ( e^{\sigma}(R' \lambda' - 2 R'' - R' \sigma') - e^{\lambda}(\dot{R}\dot{\sigma} -  2\ddot{R} - \dot{R} \dot{\lambda} )) \nonumber
    \\ &&
    =  (\tilde{\kappa} (\rho - p) + 2\Lambda) R^2 e^{\lambda + \sigma} \nonumber
    \end{eqnarray}
\bibliographystyle{unsrt}
\bibliography{references}

\end{document}